\documentclass[conference,a4paper]{IEEEtran}
\IEEEoverridecommandlockouts
\usepackage{cite}
\usepackage{amsmath,amssymb,amsfonts}
\usepackage{soul}
\usepackage{multicol}
\usepackage{algorithm}
\usepackage{algpseudocode}
\usepackage{booktabs}
\usepackage{datetime}
\usepackage{graphicx}
\usepackage{textcomp}
\usepackage[draft]{hyperref}
\usepackage{xcolor}
\usepackage{tabularx}
\usepackage{todonotes}
\usepackage{standalone}
\usepackage{balance}
\usepackage[inkscapelatex=false]{svg}
\usepackage{url}
\def\BibTeX{{\rm B\kern-.05em{\sc i\kern-.025em b}\kern-.08em
    T\kern-.1667em\lower.7ex\hbox{E}\kern-.125emX}}
\usepackage{siunitx}

\newcommand{\byte}{B}

\renewcommand\IEEEkeywordsname{Keywords}

\algdef{SE}[CALLBACK]{Callback}{EndCallback}
   {\algorithmiccallback}
   {\algorithmicend\ \algorithmiccallback}
\algnewcommand{\algorithmiccallback}{\textbf{callback} }

\algdef{SE}[ISR]{ISR}{EndISR}
   {\algorithmicisr}
   {\algorithmicend\ \algorithmicisr}
\algnewcommand{\algorithmicisr}{\textbf{isr} }

\newcommand{\MYheader}{2022 International Conference on Indoor Positioning and Indoor Navigation (IPIN), 5 – 7 Sep. 2022, Beijing, China}
\newcommand{\MYcopyrigth}{978-1-7281-6218-8/22/\$31.00~\copyright~2022~IEEE\hfill}

\renewcommand{\MYheader}{}
\renewcommand{\MYcopyrigth}{}

\makeatletter
\def\ps@headings{%
	\def\@oddhead{\hfill\MYheader\hfill }
	\def\@evenhead{\hfill\MYheader\hfill}
	\def\@oddfoot{}%
	\def\@evenfoot{}}
\def\ps@IEEEtitlepagestyle{%
	\def\@oddhead{\hfill\MYheader\hfill}
	\def\@evenhead{\hfill\MYheader\hfill}
	\def\@oddfoot{\MYcopyrigth}%
	\def\@evenfoot{\MYcopyrigth}
    }
\makeatother
\begin{document}
\title{What Your Wearable Devices Revealed About You and Possibilities of Non-Cooperative 802.11 Presence Detection During Your Last IPIN Visit\\
\thanks{The authors gratefully acknowledge funding from European Union’s Horizon 2020 Research and Innovation programme under the Marie Sk\l{}odowska Curie grant agreement No. $813278$ (A-WEAR: A network for dynamic wearable applications with privacy constraints, \url{http://www.a-wear.eu/}) and No. $101023072$ (ORIENTATE: Low-cost Reliable Indoor Positioning in Smart Factories, \url{http://orientate.dsi.uminho.pt/}). This work does not represent the opinion of the European Union, and the European Union is not responsible for any use that might be made of its content.}}

\author{\IEEEauthorblockN{Tomáš Bravenec\IEEEauthorrefmark{1}\IEEEauthorrefmark{2},
Joaquín Torres-Sospedra\IEEEauthorrefmark{3}, Michael Gould\IEEEauthorrefmark{1} and Tomas Fryza\IEEEauthorrefmark{2}}
\IEEEauthorrefmark{1}\textit{Institute of New Imaging Technologies, Universitat Jaume I}, Castellón, Spain \\
\IEEEauthorrefmark{2}\textit{Department of Radio Electronics, Brno University of Technology}, Brno, Czech Republic \\
\IEEEauthorrefmark{3}\textit{Algoritmi Research Centre, University of Minho}, Guimarães, Portugal\\ \texttt{bravenec@uji.es} -- \texttt{jtorres@algoritmi.uminho.pt} -- \texttt{gould@uji.es} -- \texttt{fryza@vut.cz}}

\maketitle

\begin{abstract}
The focus on privacy-related measures regarding wireless networks grew in last couple of years. This is especially important with technologies like Wi-Fi or Bluetooth, which are all around us and our smartphones use them not just for connection to the internet or other devices, but for localization purposes as well. In this paper, we analyze and evaluate probe request frames of 802.11 wireless protocol captured during the 11\textsuperscript{th} international conference on Indoor Positioning and Indoor Navigation (IPIN)~2021. We explore the temporal occupancy of the conference space during four days of the conference as well as non-cooperatively track the presence of devices in the proximity of the session rooms using 802.11 management frames, with and without using MAC address randomization. We carried out this analysis without trying to identify/reveal the identity of the users or in any way reverse the MAC address randomization. As a result of the analysis, we detected that there are still many devices not adopting MAC randomization, because either it is not implemented, or users disabled it. In addition, many devices can be easily tracked despite employing MAC randomization. 
\end{abstract}

\begin{IEEEkeywords}
MAC randomization, temporal analysis, privacy, probe requests
\end{IEEEkeywords}

\section{Introduction}

During the last decade, the development of new wireless technologies continued and with it the field of indoor positioning and indoor navigation. Positioning and navigation indoors are more difficult than outdoors, where Global Navigation Satellite Systems (GNSSs) are widely adopted. That is due to the inability of outdoor positioning system signals to penetrate the walls of buildings. In addition, the heterogeneity of indoor spaces makes positioning even more challenging. This means the systems for indoor positioning and navigation require different technologies. Since the technologies suitable for these applications are already deployed (Wi-Fi) or the beacons are easy to deploy (Bluetooth). With these, the issues of privacy come to the surface.

User privacy with wearable devices is a big topic, as it is difficult to find a balance between enough privacy and functionality. That is due to many services requiring user data to provide useful and helpful information. That is especially true in positioning applications, in which the position is not calculated on the device. That is due to the fact that the server needs some identifier to send the location data back as well as the signal strengths or channel state information and other information useful for localization. This information is known only to the server and is therefore transmitted using encrypted communications. 

In this work, we focus only on the Wi-Fi communication protocol and its unencrypted management frames. These frames are not only used for managing the connection to an access point, but also for detection of nearby access points using probe request frames. The detected networks are then used either for connection to known networks saved in the preferred network list of the device or for approximate localization without global positioning systems. A side effect of using these scans for nearby networks is the ability of adversaries to collect these unencrypted frames.

The main paper contribution is a look at presence detection and user tracking at an isolated conference using not encrypted management frames of Wi-Fi protocol. We show the simplicity of tracking devices not employing MAC address randomization, as well as some devices that do. On the contrary, we also present the appearance of devices with well-implemented MAC address randomization. 

In this paper we first provide an overview of related works in Section~\ref{section:related}, followed by Section~\ref{section:current_mac_randomization} with description of current implementation of MAC address randomization depending on platforms. In Section~\ref{section:dataset} in which we first provide information on the ethics of the capture of user data, which we follow with the description of the conference space. Furthermore, we describe the creation of the IPIN 2021 Probe Request dataset. Section~\ref{section:analysis} presents the results of the analysis of the created dataset, with several points of view on the gathered data. In the end, in  Section~\ref{section:conclusions} we discuss the results, conclude the paper, and provide the lines for future work.


\section{Related Work} 
\label{section:related}

The analysis of privacy weaknesses in management frames of 802.11 protocol is not new and has been explored for user tracking in the past. Out of the management frames, probe requests are most vulnerable to tracking. 

In 2012 the authors of~\cite{tracking_2012_paper} exploited the availability of unique identifiers (MAC address) in the probe request for urban mobility tracking. Before the introduction of privacy measures considering Wi-Fi probe requests, specifically MAC address randomization in 2014 by Apple in the operating system iOS~8~\cite{ios8_hutchinson_2014}, the Sapienza Probe Request Dataset was published~\cite{sapienza-probe-requests-PAPER, sapienza-probe-requests-CRAWDAD}. Several researchers already proved the vulnerability of probe requests before the implementation of MAC address randomization in~\cite{cunche2014linking, cheng2013characterizing}. 

Since the introduction of MAC address randomization, researchers focused on exploring ways to bypass the newly introduced privacy measures and revealing the globally unique MAC address of each device using locally assigned MAC address~\cite{mac_decomposition_2016}. One year after the introduction of MAC address randomization, researchers worked on reverse-engineering the MAC address randomization Apple had used~\cite{talkative_device_2015_paper}. Researchers also captured probe requests of Wi-Fi users in Italy during two political events and focused on figuring out the origin of the participants~\cite{di2016mindYourProbes}. Their results were very closely matching the official voting reports. Other forms of analysis used temporal differences between subsequently transmitted probe requests to distinguish different devices~\cite{Matte_timing_2016}. The authors of~\cite{Mac_rand_when_fail_2017} compiled a very comprehensive study of privacy-related measures in probe request frames of 802.11. This study did explore when exactly the MAC address randomization is not enough and when exactly it fails to protect the user privacy. In 2020 the authors of~\cite{probe_based_identification_2020} successfully explored ways of protecting user privacy by encrypting probe requests, but the encryption of probe request frames was not adopted by the industry yet. As a follow-up to~\cite{Mac_rand_when_fail_2017}, another deep study of MAC randomization was published~\cite{3_years_later_mac_succeeds_2021}, which provides a deep look into the progress of protecting user privacy over the years.

Industry introduced the MAC address randomization in \num{2014}~\cite{ios8_hutchinson_2014}, but its implementation is still not perfect as sensitive information (enabling user's tracking) is still leaked. 
To explore the vulnerability of the implementations of MAC randomization we decided to use an environment of a scientific conference. The contribution of this paper is in the analysis of the current state of privacy-related measures in Wi-Fi probe requests around an isolated scientific event and the possibilities of non-consensual presence detection. 


\section{Current MAC Randomization Implementation}
\label{section:current_mac_randomization}

Implementation of randomizing MAC addresses is varying depending on the manufacturers and software developers. This fragmentation in implementation exists due to the lack of a commonly followed standard for MAC randomization. Few years after the MAC randomization was introduced, the specification of a standard amendment 802.11aq-2018~\cite{ieee802.11aq_standard} was specified by IEEE SA Standards Board in \num{2018}, but the implementation itself differs between manufacturers. 

\subsection{Identification of Randomized MAC Address}

\begin{figure}[!hbt]
    \centering
    \includegraphics[width=\hsize]{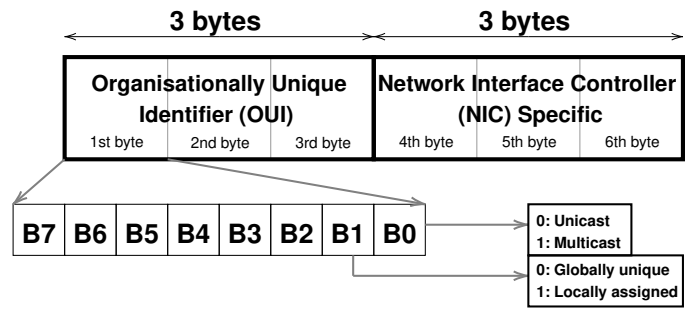}
    \caption{Structure of MAC address with the functional bits}
    \label{fig:mac_address_structure}
\end{figure}

Even with the fragmentation in implementations, all implementations follow the setup of the two least significant bits in the first byte of MAC address as shown in Fig.~\ref{fig:mac_address_structure}. In case the 2nd least significant bit of first byte B1 is set, the MAC address was assigned locally by the network controller of the device. The least significant bit of the first byte B0 distinguishes between individual devices and device groups. Since randomized MAC address will always have bit B1 set to \num{1} and individual devices have bit B0 cleared to \num{0}, recognition of randomized MAC address is simple. Due to the fixed values the two least significant bits can have, the 2nd digit of randomized MAC address in hexadecimal format has only four options: \num{2} (0010), \num{6} (0110), A (1010) or E (1110).

\subsection{Google Android}
\label{section:android_implementation}

The MAC address randomization was supported by Android since version~\num{6}, although the implementation of randomized MAC addresses for probing was first included with Android~8~\cite{android_rnd_implementation}. Android~\num{9} first implemented an option to connect to a Wi-Fi network using a randomized address, even though it was disabled by default and only available through developer options. Starting with Android~\num{10}, MAC randomization is enabled by default~\cite{android_rnd_implementation} for every new network and randomizes the MAC address for every SSID. The randomized address does not change as the generated address depends on the network profile (SSID, security and so on), non-persistent randomization that might change the MAC address for every connection to a network can also be enabled in the developer options~\cite{android_rnd_behaviour}.

In the first versions of Android supporting MAC address randomization, the system randomized only the last \num{3} bytes of the address, while using a fixed prefix for the first \num{3} bytes. Android devices using this older MAC randomization implementation had their randomized MAC address starting with DA:A1:19 prefix~\cite{Mac_rand_when_fail_2017}.

There is one catch to this though, even though the latest version is Android~\num{12} at the time of writing (May 2022), barely any device is actually running it. Even though the majority of devices are running Android~\num{10} or higher, there are still a lot of devices running older versions of Android, which do not use MAC randomization or any other privacy measures when it comes to probe requests. 

\subsection{Microsoft Windows}
\label{section:windows_implementation}
 
Similar to Google's Android, the operating system developed by Microsoft supports MAC address randomization per SSID. The MAC randomization is present in Microsoft's operating system since version~\num{10}~\cite{huitema_W10_2015} and is turned on by default. The default option uses the same randomized MAC address for each SSID, and the MAC address stays randomized for the actual connection to the access point. This approach is the same as in Android and helps keep MAC address-based authentication working. On the other hand, there is an option to change the MAC address daily, but this needs to be enabled manually as this approach could cause issues with connection to some networks using MAC address authentication. The implementation of MAC address randomization in the newest version of the system developed by Microsoft stayed the same as in Windows~\num{10}, with MAC addresses staying the same for one SSID. 

\subsection{Apple iOS}
\label{section:ios_implementation}
 
Apple first introduced MAC address randomization with iOS~\num{8} in \num{2014}~\cite{ios8_hutchinson_2014}. Then later on iOS~\num{10} added a tag in the information element of the probe request, which allowed for simple identification of iOS devices. The last change to the implementation of MAC address randomization happened in iOS~\num{14}. Since iOS~\num{14} the devices use randomized MAC address per SSID just like Microsoft Windows and Android devices running the latest versions of their own systems. The implementation of MAC address randomization on iOS~\num{15}~\cite{apple_ios_2021} changed a little. The modifications were related to the changes in the locally assigned address of the device for one network. This happens on multiple occasions:

\begin{itemize}
    \item on forgetting the network and reconnecting again,
    \item if the device did not connect to the network for 6 weeks,
    \item on-device content reset or network settings reset.
\end{itemize}

This all makes the implementation of MAC address randomization on Apple devices very robust, while not breaking existing systems with MAC address authentication.

\subsection{Other devices}

From other common operating systems, the implementation of MAC addresses again differs on the developer or the manufacturer of the device. Linux supports MAC address randomization since \num{2014}, specifically kernel 3.18.~\cite{linux_318_MAC_random_commit}. For Apple devices, the support changes from device to device. 
The support for the non-iOS devices produced by Apple varies based on the generation of the product~\cite{apple_rand}. 

\section{Dataset Creation}
\label{section:dataset}

Since the investigation into probe requests related to the IPIN conference required capturing a new dataset during the conference, we wrote a firmware for a Wi-Fi enabled microcontroller and used it to collect the probe requests transmitted in the \SI{2.4}{\giga\hertz} frequency band. The probe request sniffer was active starting on 29~November, \formattime{8}{22}{0} until the end of the closing ceremony on 2~December, \formattime{13}{02}{0}. During this time, we captured a total of \num{390810} probe requests.

\subsection{Probe Request Sniffer}

The device we used for collecting probe requests is an ESP32 microcontroller with custom firmware available from Gitlab repository~\cite{probeSniffer_2021}. 

Since probe requests do not contain time information, the ESP32 first connects to a predefined Wi-Fi to download current time information. After getting the current time, the SD card is connected, and the wireless interface is switched to monitoring mode. 

While in monitoring mode, every collected frame is checked for its type. In case of capturing a probe request, the frame is stored in a file on the SD card. All other frames are discarded and therefore, not recorded. The file is periodically saved and a new one gets to be created to prevent data corruption in case of power loss. 

The sniffing of probe requests continues until the \textit{Stop} button is pressed, which raises an interrupt which stops the data collection and safely disconnects the SD card. The firmware of the ESP32 is simplified in Algorithm~\ref{alg:esp_firmware}.

\begin{algorithm}    
    \caption{Probe Request Sniffer}
    \label{alg:esp_firmware}
    \begin{algorithmic}[1] 
        \State Initialize MCU peripherals and GPIO
        \State Connect to Wi-Fi and download current time
        \State Initialize and mount SD card
        \State Check for existing files and open new pcap file
        \State Reinitialize Wi-Fi in monitoring mode
        \State Start probe sniffing task - run \textbf{callback} Received Packet
        \State Start saving task - periodically run \textbf{callback} Save PCAP
        \Statex
        \Callback {Received Packet}
            \If {packet.type = ProbeRequest}
                \State Write packet to file
            \EndIf
        \EndCallback
        \Statex
        \Callback {Save PCAP}
            \State Close current pcap file
            \State Open new pcap file
        \EndCallback
        \Statex
        \ISR {On Button Press}
            \State Stop probe sniffing task
            \State Close pcap file and unmount SD card
        \EndISR
    \end{algorithmic}
\end{algorithm}

The sniffer is detectable by other wireless devices only during the time acquisition period as it involves bi-directional communications. After downloading the current time, the Wi-Fi interface of the device is set to ``monitoring only''. The sniffer is collecting 802.11 probe request frames without being detectable as the wireless interface is not transmitting any packets. i.e., it is only passively receiving packets from the radio environment around it.

\subsection{Ethics and Sensitive Information}

One thing to mention about the captured data is the fact, that right after capture by the ESP32-based probe request sniffer, it does contain the user information. That is to reduce the computational complexity of the probe request sniffer and produce almost the same packet capture files as network analysis tools like Wireshark. The captured data are then exactly the same as transmitted by devices and contain sensitive information. Be it globally unique or randomized MAC addresses, leaking SSIDs from devices preferred networks list, up to the device manufacturers and device names. 

Additionally, since the captured probe requests are from in-person and a quite isolated event, with minimal presence of people not participating in the IPIN conference, it is necessary to say, that first of all we anonymized the captured data in a~way we could not get any personal information during the analysis itself. The anonymization was done by hashing fields containing personal information with SHA512. Using hashing algorithms on the user information ensures the anonymity of the users while preserving the option for analysis. The analysis was focused on the vulnerability of the management frames of the current implementation in the 802.11 protocol, not on linking private information to specific users.

Even though the captured data does contain real globally unique MAC addresses and many randomized ones, thanks to the anonymization we did before the analysis itself, it is not possible to link the specific individuals to a MAC address or identify anyone directly. The anonymized version of the dataset is publicly available from Zenodo repozitory~\cite{bravenec_tomas_2022_6798302}.

\subsection{Space Description}

The conference took place in Lloret de Mar, Spain in Evenia Olympic Congress Centre from 29~November to 2~December. The only people present around the hotel lobby and near the session rooms from the beginning to the end of the conference were attendants of the conference, conference organizers, hotel employees and cleaning staff.

\begin{figure}
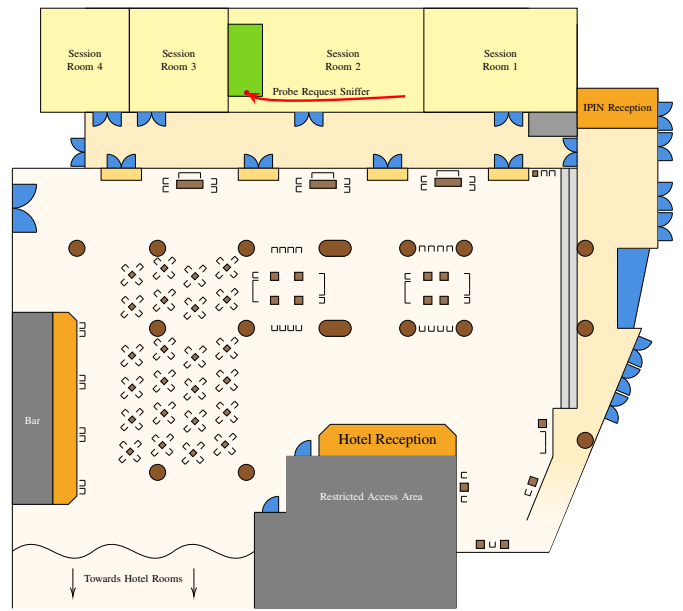

    \centering
    \includestandalone[width=\hsize]{Images/ipin_map}
    \caption{Floorplan of the Evenia Olympic Congress Centre (Lloret de Mar, Spain)}
    \label{fig:conference_map}
\end{figure}

The entire conference space was around the lobby, with hotel rooms and hotel restaurants being far enough to not pose interference and capture probe requests from sources we did not care about, similarly the location of the conference was in a single-floor section of the hotel complex, which guaranteed that all of the collected probe requests were from the area of the actual conference. The probe request sniffer was placed under the stage in Session Room 2 as presented in Fig.~\ref{fig:conference_map}. The entire conference space was in the radio range of the sniffer. 

\subsection{Data Description}

During the four days of the conference, we captured \num{390810} probe requests. The capture started 38 minutes before the first tutorial session, on \formatdate{29}{11}{2021} at \formattime{8}{22}{40}. The last probe request was then captured on \formatdate{2}{12}{2021} at \formattime{13}{02}{49}, just few minutes after the closing ceremony concluded. Unfortunately, we were unable to keep the ESP32 working past the closing ceremony, due to the preparation of the session rooms for the next event.

The probe requests captured during the conference contained many of the additional fields carrying extra information about the capabilities of the origin device. Be it information about supported data rates, capabilities related to certain Wi-Fi standards (HT and VHT Capabilities), vendor-specific elements or Wi-Fi Protected Setup fields with UUID-E. All of the fields used for creating the fingerprint of each device are in Table~\ref{tab:probe_fields}.

\begin{table}[!hbt]
    \centering
    \caption{Probe request fields used to create device fingerprint and frequency of occurrence in data collected in our lab}
    \label{tab:probe_fields}
    \begin{tabularx}{\columnwidth}{llS[table-format=6.0]S[table-format=3.2]}
    \toprule     
        \multicolumn{2}{c}{Information Element}            & {Included in Probes}      & [\SI{}{\percent}] \\ \midrule
        \multicolumn{2}{l}{Supported rates}                & 390211                    & 99.85             \\
        \multicolumn{2}{l}{Extended Supported rates}       & 385606                    & 98.67             \\
        \multicolumn{2}{l}{HT Capabilities}                & 359391                    & 91.96             \\
        \multicolumn{2}{l}{VHT Capabilities}               &  51031                    & 13.06             \\
        \multicolumn{2}{l}{Extended Capabilities}          & 312181                    & 79.88             \\
        \multicolumn{2}{l}{Vendor Specific elements}       & 228970                    & 58.59             \\
        & 1 Vendor Specific element                        &  84215                    & 21.55             \\
        & 2 Vendor Specific elements                       &  67663                    & 17.31             \\
        & 3 Vendor Specific elements                       &  55524                    & 14.21             \\
        & 4 Vendor Specific elements                       &  21462                    &  5.49             \\
        & 5+ Vendor Specific elements                      &    106                    &  0.03             \\
        \multicolumn{2}{l}{WPS - UUID-E}                   &   3733                    &  0.96             \\
        \multicolumn{2}{l}{WEP Protected}                  &    599                    &  0.15             \\ \midrule
        \multicolumn{2}{l}{Total Collected Probe Requests} & \multicolumn{2}{S[table-format=12.0]}{390810} \\ \bottomrule
    \end{tabularx}
\end{table}


\begin{figure*}
    \centering
    \includegraphics[width=0.95\hsize]{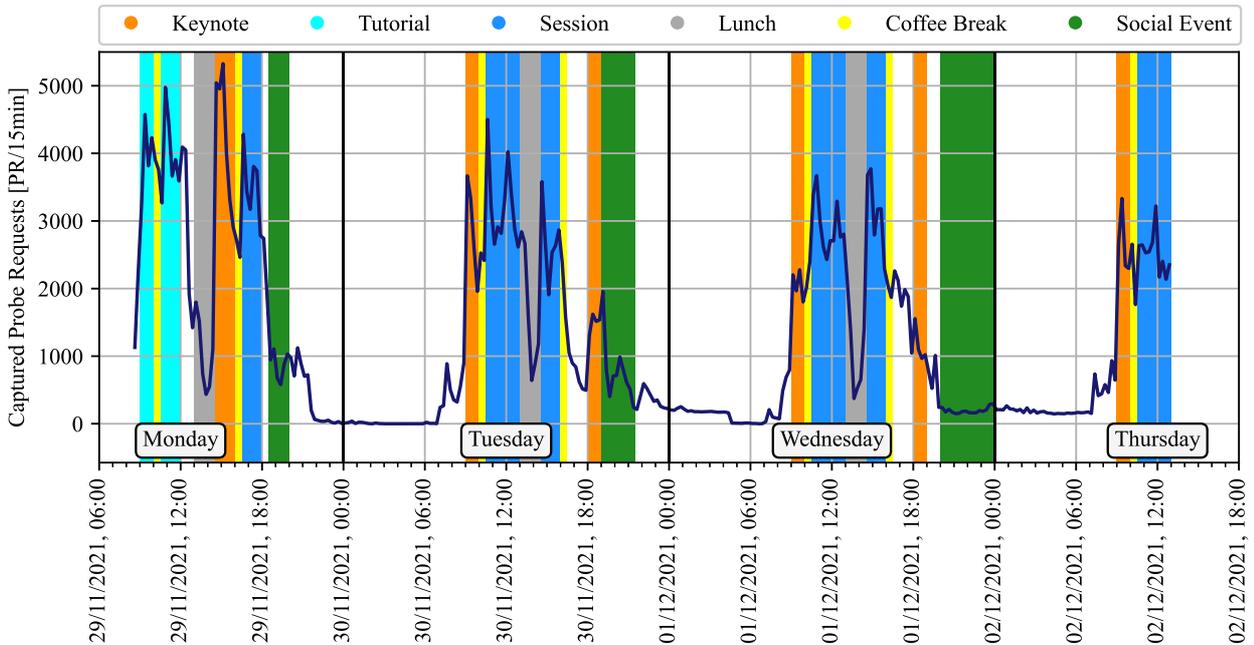}
    \caption{Density of captured Probe Requests correlated with the program of the conference (amount of probe requests grouped in 15-minute clusters)}
    \label{fig:probe_density}
\end{figure*}

During the conference, we also captured more unusual types of probe requests, which did not contain any information other than \SI{22}{\byte} long encrypted data sequence, which was identical in all \num{599} occurrences. These encrypted probe requests also in \num{44}/\num{599} cases contained randomized MAC address.

\section{Analysis \& Results}
\label{section:analysis}

To analyze the gathered probe requests, we looked at the data from several perspectives. Using the gathered probe requests for crowd presence detection. Then we analyze the impact of MAC randomization on our ability to track individual users, both with MAC randomization on and off.

\subsection{Presence detection}
\label{section:presence}

We counted the amount of probe requests received every couple of minutes, to see how big of a crowd could we detect from the density of probe requests. We chose a 15-minute interval as a compromise between readability and resolution of the final plot. From Fig.~\ref{fig:probe_density} we can see that during every keynote, tutorial or session the presence of users was much higher compared to the breaks in between the sessions. This can be caused by people turning off computers for the duration of coffee breaks. Quite a lot of people also left the range of the sniffer to go into the hotel restaurants for lunch. During the nighttime we can also see, if someone stayed in the lobby, be it for work or socializing. During the night from Monday to Tuesday, the lobby stayed empty with someone going through it but not staying long. The next two nights, the lobby was not empty during the night.

From the data plotted in Fig.~\ref{fig:probe_density} we can also see which keynote or session group (IPIN 2021 had 4 parallel session tracks) was more interesting to the participants of the conference. Unfortunately, due to the deployment of only one Probe Request Sniffer, we could not implement an indoor localization method based on RSSI to determine which session room the participants of the conference occupied at any time.

The Tuesday social event (Networking in the Kitchens) took place mostly out of the range of the probe sniffer in one of the hotel's restaurants. After the event, some of the participants stayed for further socializing, which can be seen on the small local peak right after the event ended. Another drop in received probes happened on Wednesday during the gala dinner, which took place in neighboring village and the presence in the conference space was minimal.

One of the noticeable trends is also the drop in the amount of captured probe requests during coffee breaks. This indicates people leaving the area either to get some fresh air outside of the hotel lobby, use the restroom or go to their hotel rooms. Since the amount of probe requests increases after each coffee break again, it is safe to assume that the conference attendants were coming back after each of the coffee breaks was over. 

\subsection{Analysis of user presence with global MAC address}

\begin{figure}[!hbt]
    \centering
    \includegraphics[width=\hsize]{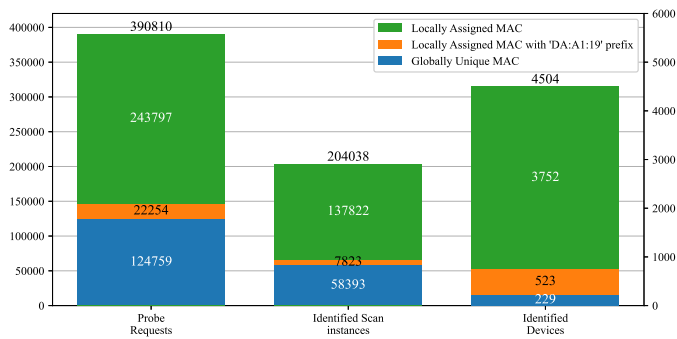}
    \caption{Randomized MAC addresses in probe requests, identified scan instances and distinguished devices}
    \label{fig:bar}
\end{figure}

It is no surprise that devices that transmit their real MAC address are very easy to track. Since we are able to identify probe requests using their globally unique identifier, their identification is very simple. At the IPIN 2021 conference, \SI{28.62}{\percent} of identified scan instances (\num{58393} of \num{204038} scan instances) used their globally unique MAC address. Since these devices used their real MAC address, we clustered the identified instances together and distinguished \num{229} individual devices without MAC address randomization through the duration of the conference. This data can be seen in Fig.~\ref{fig:bar}. We then explored the presence of these devices in the proximity of our probe request sniffer. This presence in time proved to us how easy it is to track devices that do not employ any privacy-related measures related to the probe requests. The temporal presence of \num{10} devices using their real MAC address in the conference space is in Fig.~\ref{fig:temporal_global_mac} as an example of the simplicity of tracking these devices.

\subsection{Analysis of user presence with local MAC address}

On the contrary, tracking devices which do not transmit unique identifiers is much more challenging. Out of the captured probe requests, \SI{68.08}{\percent} (\num{266051}) were using locally assigned MAC addresses. Since the devices do not change the MAC address they transmit during the Wi-Fi search burst, we were able to identify \num{204038} different scan instances using just the MAC addresses available from the captured probe requests.

Out of all the transmitted probe requests with randomized MAC address, \num{22254} used \textit{DA:1A:19} as the first \num{3} bytes of their MAC address. We identified \num{7823} individual scan instances using this prefix. After we matched these instances together using fingerprinting of information elements, the similarity between transmitted Preferred Network Lists and the recurrence of the same randomized MAC addresses, we identified \num{523} devices using the \textit{DA:1A:19} MAC address prefix. These data are presented in Fig.~\ref{fig:bar} with the comparison to the number of devices with a fully randomized MAC address and with a globally unique one. \num{50} of these devices then showed up more than \SI{10}{\times} (we chose \num{10} as a threshold as we found out that devices with more than \num{10} appearances are easy to track over time).

After identifying individual scan instances, we used the same approach consisting of fingerprinting, the similarity of preferred network lists and reappearance of MAC addresses to match together all other devices as well. With this approach we initially distinguished \num{4274} devices using locally assigned MAC address out of which \num{3752} randomized \num{46} out of the \num{48} bits in a MAC address. In this initial number of devices, \num{3544} appeared less than \num{10} times. On the other hand, \num{296} devices with fully randomized MAC address showed up more than \SI{10}{\times} which made them easily identifiable despite them using randomized MAC address, as can be seen from \num{10} example devices in Fig.~\ref{fig:temporal_local_mac}.

\subsection{Single Occurrence of Devices in Time Domain}

\begin{figure}[!hbt]
    \centering
    \includegraphics[width=0.94\hsize]{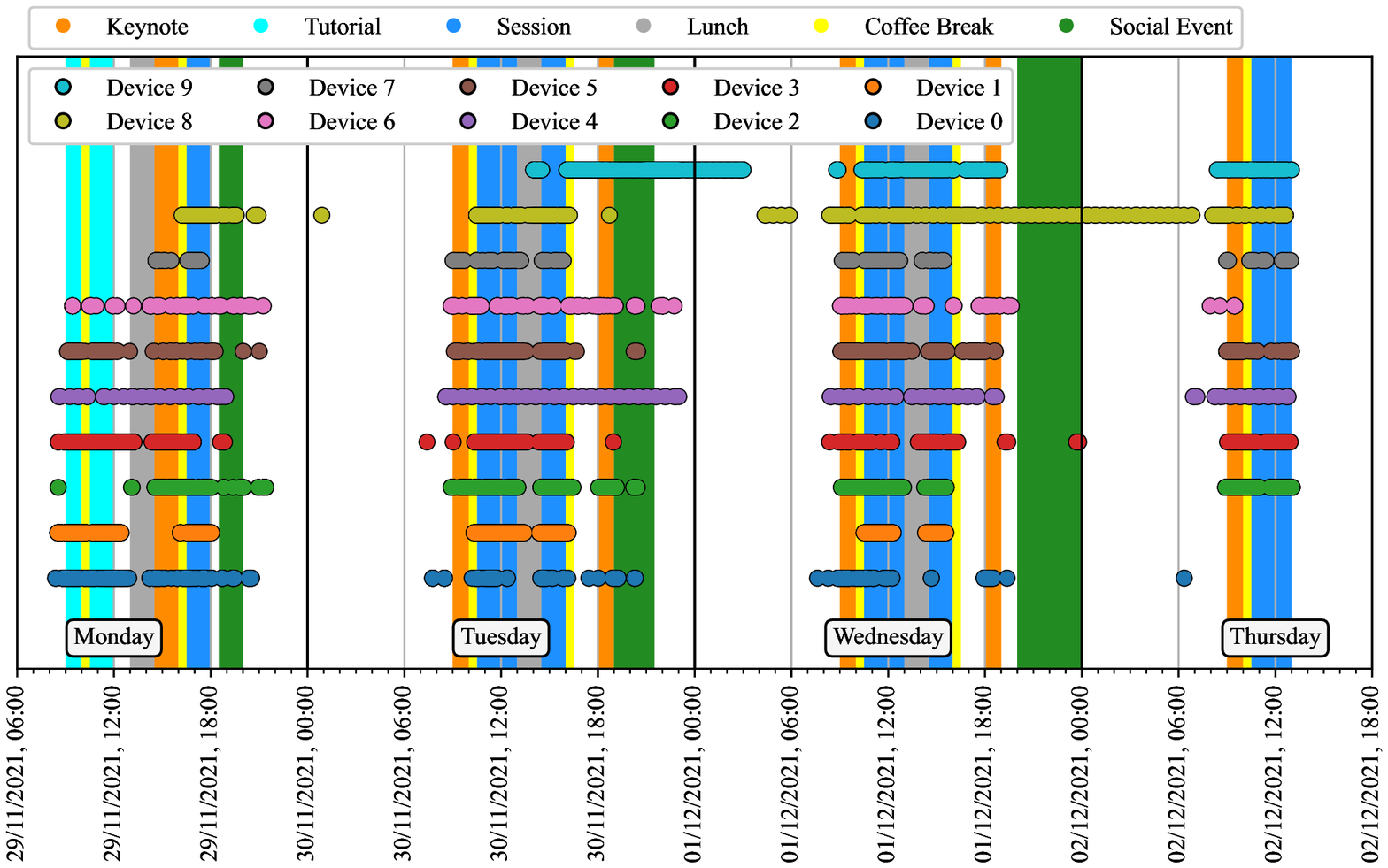}
    \caption{Repeated occurrences of devices identified by the usage of globally unique MAC address}
    \label{fig:temporal_global_mac}
    \vspace{12pt}
    \includegraphics[width=0.94\hsize]{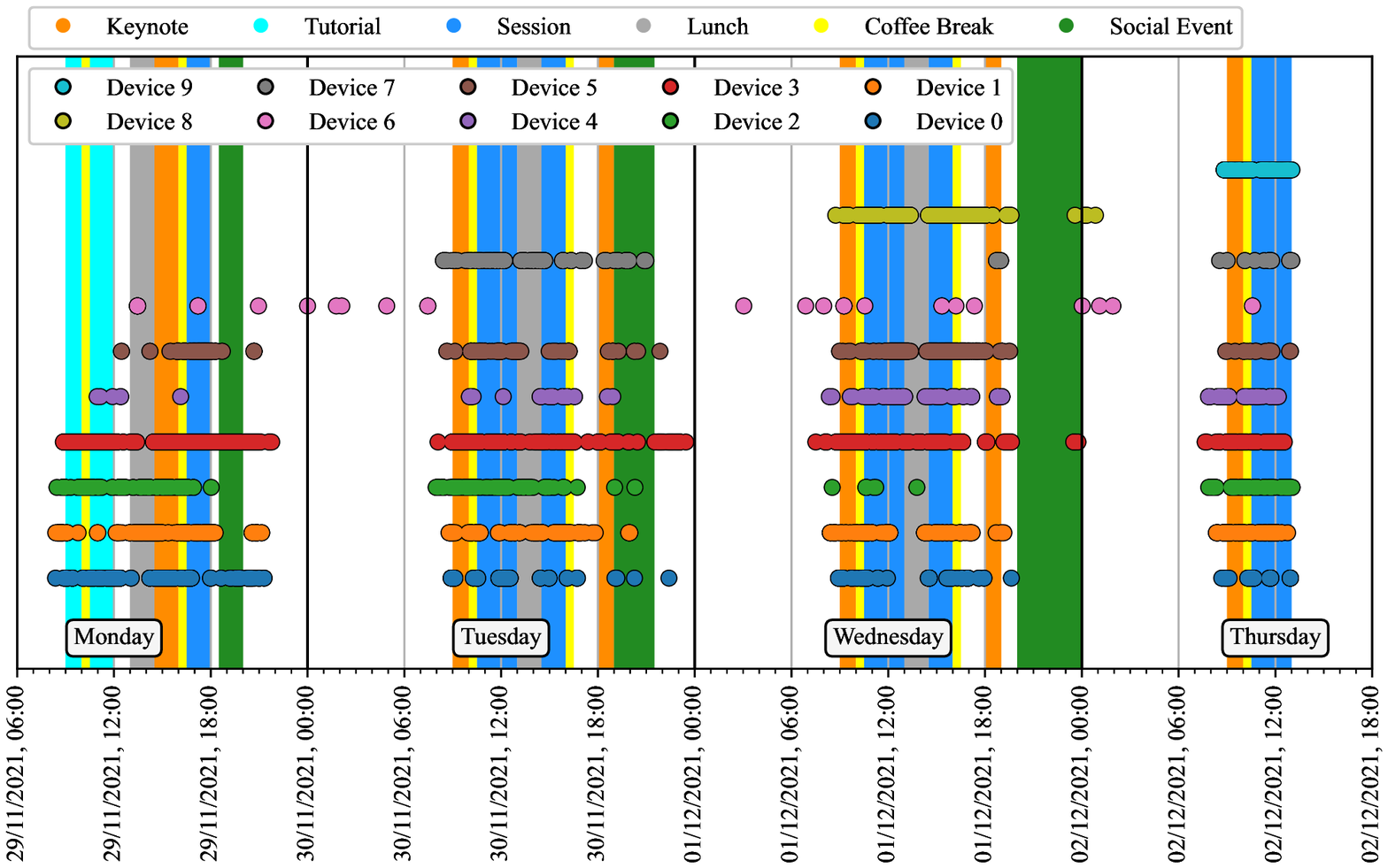}
    \caption{Recurrent identification of the same devices despite using locally assigned MAC address}
    \label{fig:temporal_local_mac}
    \vspace{12pt}
    \includegraphics[width=0.94\hsize]{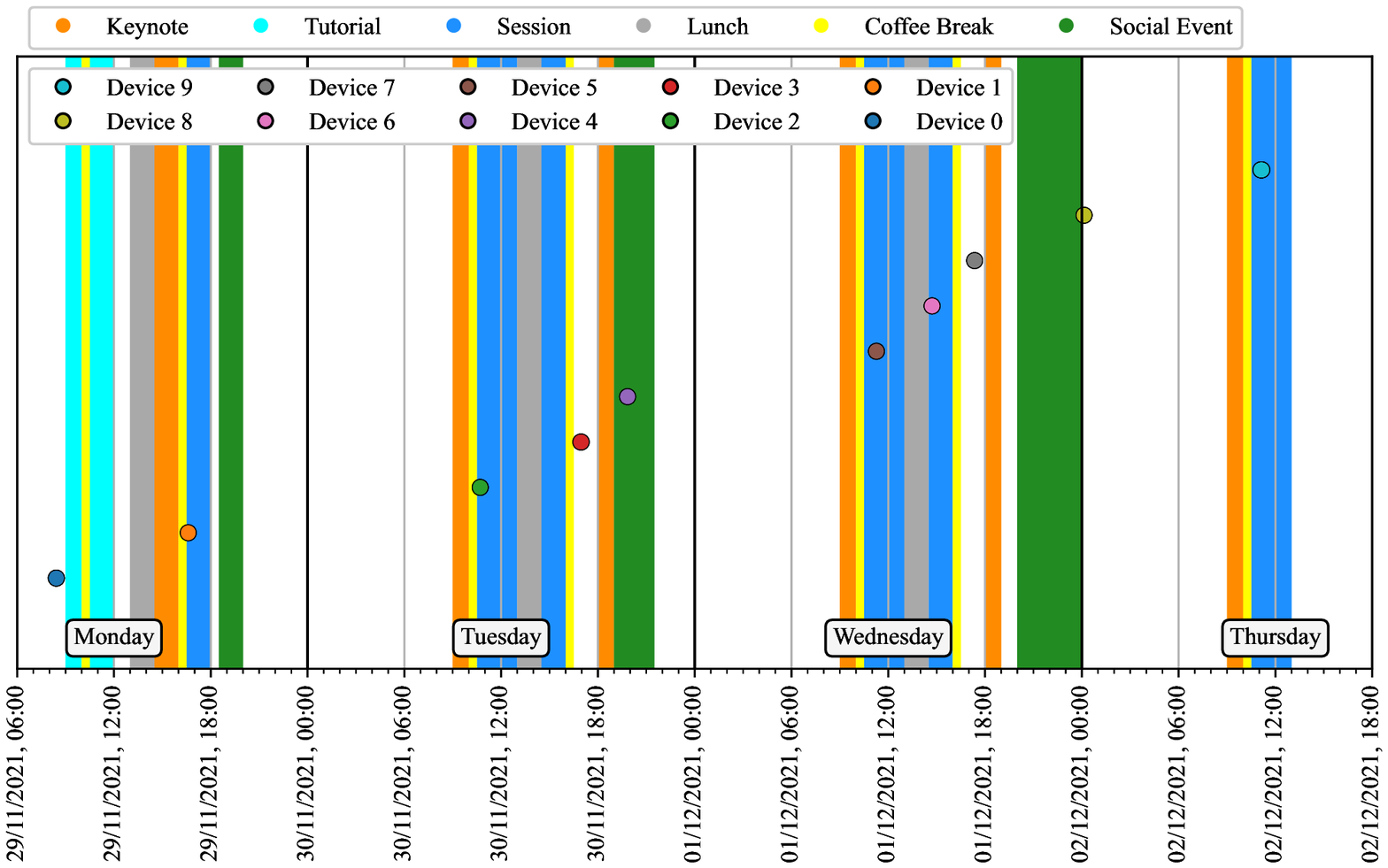}
    \caption{Detected devices without identified recurrences in time}
    \label{fig:single_appearance}
\end{figure}

From Fig.~\ref{fig:bar} we can see that the number of identified devices is still really high for just \num{3} full days in a conference space. Especially since the event space was primarily occupied by the attendants of the conference and hotel staff. The sniffer was also in the range of the sidewalk next to the entrance of the hotel. It is quite possible many of the single occurrences were just from pedestrians walking in the proximity of the sniffer. Another reason for this is a good implementation of MAC address randomization, the transmission of reduced information elements in the probe requests and omitting the transfer of SSIDs from the saved preferred network list. Representation of unmatched devices is shown in Fig.~\ref{fig:single_appearance} with \num{10} examples. 

\section{Conclusions and Future Work}
\label{section:conclusions}

In this paper, we explored the possibilities of passive presence detection and tracking at~\num{4} days-long scientific conference. We did this by exploiting unencrypted management packets of the 802.11 communication protocol. Specifically, we focused on probe request frames.

We used the ESP32 microcontroller with custom firmware for sniffing Wi-Fi probe requests. After the data collection, we analyzed the data and matched together devices that had the same MAC address, be it a globally unique one or a locally assigned one. We also used information elements fingerprinting and similarity of proffered network lists to further identify different scan instances as one device. Not surprisingly, the devices without a randomized MAC address were easily tracked. As we expected, devices employing MAC address randomization were more difficult to track, but even then, we identified \num{296} devices reappearing in time. 

Even though the manufacturers are employing privacy-related measures like MAC address randomization since \num{2014}, many devices are still easily tracked. That is not to say that MAC address randomization is not working, as we have seen with many appearances of only \num{1} or \num{2} scan instances. For these devices, the implementation of locally assigning MAC address either works well or we were also capturing probe requests from the pedestrians using the sidewalk next to the hotel lobby. In any case, we expect the situation to get better in time with older devices (with worse implementations of MAC address randomization) being replaced by newer devices after reaching their end of life.

To continue this research work, we are going to further explore privacy-related measures in wireless networks. We plan to study and search for user information leaks in Wi-Fi and other wireless technologies. Another point of interest is to extend this work with passive and non-cooperative indoor localization of users. We also plan to publish the dataset used in this work as part of a new publicly available Wi-Fi probe request dataset. 

\balance

\bibliographystyle{IEEEtran}
\bibliography{main}

\end{document}